\documentclass[12pt]{article} 
\usepackage[tablesfirst,notablist,nomarkers]{endfloat} 

\usepackage{ol}
\usepackage{hyperref}
\usepackage{amsmath}

\begin{document}

\title{Quantum key distribution network with wavelength addressing}

\author{Xiao-Fan Mo, Tao Zhang, Fang-Xing Xu, Zheng-Fu Han, and Guang-Can Guo}

\address{Key Lab of Quantum Information, University of Science and Technology of China \\ Hefei Anhui 230026, China}


\begin{abstract}Most traditional applications of quantum cryptography are point-to-point communications, in which only two users can exchange keys. In this letter, we present a network scheme that enable quantum key distribution between multi-user with wavelength addressing. Considering the current state of wavelength division multiplexing technique, dozens or hundreds of users can be connected to such a network and directly exchange keys with each other. With the scheme, a 4-user demonstration network was built up and key exchanges were performed.\end{abstract} 


 ] 

\noindent Quantum cryptography is the only way to exchange keys with unconditional security through a public channel up to now. Using quantum cryptography and Vernam's cipher, perfectly secure communication is available. Since BB84, the first quantum key distribution (QKD) protocol, was introduced by Bennett and Brassard in 1984, many QKD protocols\cite{BB84,B92,Gisin} have been presented and many QKD implementations\cite{Gisin,JinJing,GermanFreespace} have been performed over fiber and free-space. Most of these implementations are point-to-point communications, in which only two users can exchange keys. Compared with modern communication systems such as Internet or cellphone network, building a QKD network that support multi-user communication is an interesting subject for physicists and engineers.

In 1994, Townsend of BT Lab presented passive optical networks, branched network and looped network.\cite{Townsend} In these networks, keys can be exchanged between the network controller and each one of N users. These networks are easy to build but inconvenient to use. A photon arrives at a random destination in a branched network. Live insertion and removal of users are not convenient in a looped network. DARPA quantum network was built up by BBN Technologies in October 2003.\cite{BBN} Using optical switches and key delay protocol, any-to-any communication is available in the network. The security of communication depends on the reliability of relay nodes.

In this letter, we present a QKD network scheme using wavelength division multiplexing. First, we describe the architecture of our network. Then we introduce operation mode of the network. At last we show experimental result of a 4-user demonstration network.

\textit{\textbf{Architecture of a wavelength addressing network.}}
In a communication network such as Internet, the destination of a data packet is recorded in the packet itself. When the packet arrives at a router or gateway, the host reads the information of destination and directs the packet to the correct path. For quantum key distribution, any active detection of a photon will cause the photon's quantum state to collapse and destroy the information carried by the photon. So our goal is to select a property of a photon as the mark of destination and design a corresponding passive optical element as a quantum router.

\begin{figure}
\centerline{\includegraphics[scale=1.2]{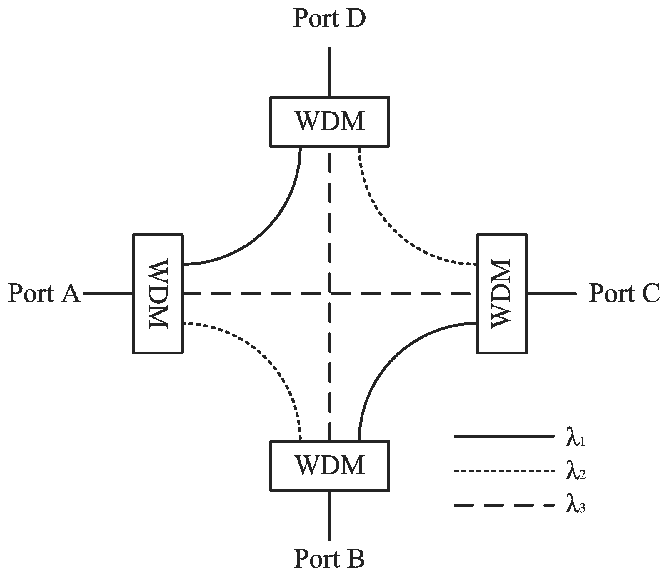}}
\caption{\label{fig:router4}Schematic diagram of 4-port quantum router: WDM, 3-channel wavelength division multiplexer; the solid lines labeled ``Port A, Port B, Port C, Port D'' are corresponding to common channels of WDMs; other lines are channels of wavelength $\lambda_1, \lambda_2$ and $\lambda_3$ of WMDs.}
\end{figure}

Here, we use a 4-user network as an example to explain the structure of a quantum router and describe how such a network works. Figure \ref{fig:router4} illustrates the structure of a 4-port quantum router, which is composed of four WDMs. Common channels of WDMs are used as I/O ports of the quantum router. Channels of wavelength $\lambda_1, \lambda_2$ and $\lambda_3$ are connected in pairs as shown in the figure. Users \textit{Alice}, \textit{Bob}, \textit{Charlie} and \textit{Delta} are respectively connected to port A, B, C and D with fiber links or free-space links. Users, optical links and the quantum router constitute a 4-user quantum network with a star topology. When \textit{Alice} transfers photons of $\lambda_1, \lambda_2$ and $\lambda_3$ to the quantum router, photons of $\lambda_1$ will be dispatched to \textit{Delta} by WMDs, photons of $\lambda_2$ to \textit{Bob}, photons of $\lambda_3$ to \textit{Charlie}. Similarly, photons of $\lambda_1$ transferred by \textit{Bob} will be dispatched to \textit{Charlie}, photons of $\lambda_2$ to \textit{Alice} and photons of $\lambda_3$ to \textit{Delta}. The complete wavelength assignment for a 4-port quantum router is listed in table \ref{tab:wavelength}. From the table, we can see that any two users of a 4-user network can directly exchange photons with three different wavelengths. When a user transfers photons of a specific wavelength to the router, the destination of photons is unique. Also, when a user detects photons of a specific wavelength, he could accurately determines from where these photons are launched. Wavelengths indicate transmission paths of photons and that is why we call such a network a wavelength addressing quantum network.

\begin{table}
\centering
\caption{\label{tab:wavelength}Wavelength assignment for a 4-port router.}
\begin{tabular}{ccccc}
	\\
	\hline
    ~      & Port A      & Port B      & Port C      & Port D      \\ \hline
    Port A & ~           & $\lambda_2$ & $\lambda_3$ & $\lambda_1$ \\
    Port B & $\lambda_2$ & ~           & $\lambda_1$ & $\lambda_3$ \\
    Port C & $\lambda_3$ & $\lambda_1$ & ~           & $\lambda_2$ \\
    Port D & $\lambda_1$ & $\lambda_3$ & $\lambda_2$ & ~           \\ \hline
\end{tabular}
\end{table}

It has been proven that, for an arbitrary $N$, we can build a $N$-port quantum router whose structure is similar to the 4-port quantum router\cite{Taozhang}. If $N$ is even, $N$ $(N$-$1)$-channel WDMs are needed to build a quantum router; if $N$ is odd, $N$ $N$-channel WDMs are needed. At present, a WDM with over one hundred channels is available in market and a WDM with over four thousand channels has been developed in laboratory\cite{WDM4200}. Hence, it is technically feasible to build a wavelength addressing quantum network of dozens or hundreds of users.

\textit{\textbf{Operation mode of a wavelength addressing network.}}
In the above 4-user network, if a user desires to transfer photons to other three users, he needs a wavelength-tunable laser diode or three fixed-wavelength laser diodes. If a user desires to detect photons, a single photon detector is needed. For all users, four wavelength-tunable laser diodes or twelve fixed-wavelength laser diodes and four single photon detectors are required. Such a configuration is flexible to use but expensive to build. Here, we introduce a configuration called server-client mode, in which a better balance between cost and functionality is achieved.

We assume that \textit{Alice} has three fixed-wavelength laser diode and each of other three users has a single photon detector. Moreover, all users is equipped with other necessary instruments for quantum key distribution. When \textit{Bob} and \textit{Charlie} are desired to exchange keys, they first send requests to \textit{Alice}. After receiving requests, \textit{Alice} starts to simultaneously transfer photons of $\lambda_2$ and $\lambda_3$ and modulates them with different phases. With standard procedure of quantum key distribution, two groups of keys are established between \textit{Alice}-\textit{Bob} and \textit{Alice}-\textit{Charlie}. Because three users use unrelated random data to modulate photons, two groups of keys are different. For example, keys of \textit{Alice}-\textit{Bob} are \{\textit{0, 1, 0, 0}\} and keys of \textit{Alice}-\textit{Charlie} are \{\textit{0, 0, 0, 1}\}. The 2nd and 4th bits of keys are not identical. \textit{Alice} uses keys of \textit{Alice}-\textit{Bob} as a reference to check which keys of \textit{Alice}-\textit{Charlie} are different. \textit{Alice} notifies \textit{Charlie} of the check result by a public channel and \textit{Charlie} inverses corresponding bits. In above example, \textit{Alice} notifies \textit{Charlie} 2nd and 4th bits are "different" and \textit{Charlie} flips 2nd bit from 0 to 1 and 4th bit from 1 to 0. After that, identical secret keys are shared among \textit{Bob} and \textit{Charlie}. We call this operation as ``key reverse operation".

We call the above configuration as ``server-client model", where \textit{Alice} plays the role of a server and other users play the role of clients. Clients send requests to a server and the server responds to requests. Server is central to the configuration and the network will not function without it. For a $N$-user network including one server, only $N$-$1$ or $N$ (depends on the parity of $N$) fixed-wavelength laser diodes and $N$-$1$ single photon detectors are required. To increase robustness and redundancy of a network, more users can be configured as servers. Another benefit of server-client model is that any number of users can simultaneously establish identical keys. Key exchange can be performed between two users (unicast), several users (multicast) or all users (broadcast). For applications such as military communications, multicast and broadcast are very useful.

\begin{figure}
\centerline{\includegraphics[scale=1.2]{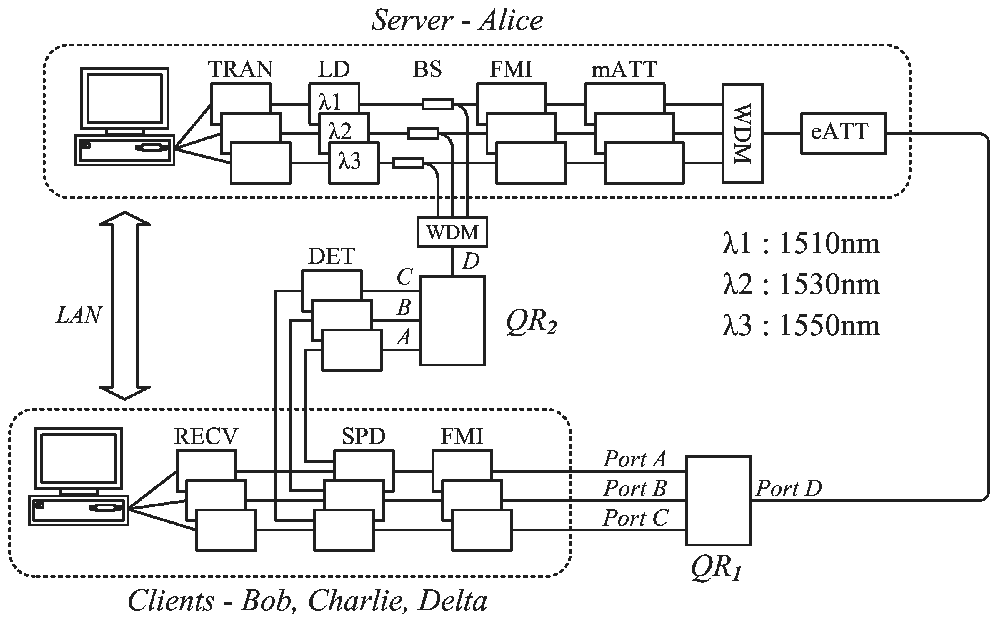}}
\caption{\label{fig:4portexp_sch}Schematic diagram of 4-user wavelength addressing network: LAN, local network; other abbreviations are deﬁned in text.}
\end{figure}

\textit{\textbf{Experiment of a 4-user wavelength addressing network.}}
With above scheme, a 4-user demonstration network was built up. Figure \ref{fig:4portexp_sch} shows the diagram of the experimental setup. Two quantum routers $\text{QR}_1, \text{QR}_2$ are used in our experiment and their structure is as shown in figure \ref{fig:router4}. Each of routers consists of four 3-channel commercial wavelength division multiplexers WDMs, where $\lambda_1$ is 1510nm, $\lambda_2$ is 1530nm and $\lambda_3$ is 1550nm. Insertion losses of different ports are listed in table \ref{tab:4portexp_insloss}. Crosstalks between different wavelength channels are about 28 to 43 dB. In our experiment, crosstalk is not essential because laser diodes are not driven simultaneously and photons of different wavelengths do not arrive at the router at same time. User \textit{Alice} is configured as a server and \textit{Bob}, \textit{Charlie} and \textit{Delta} are configured as clients. In \textit{Alice}, three transmitters are linked together by a WDM. Each transmitter consists of a laser diode LD, a Faraday-Michelson interferometer FMI\cite{JinJing}, a fixed optical attenuator mATT and a homemade circuit board TRAN. Each client is composed of a receiver, which includes a FMI, a single photon detector SPD and a homemade circuit board RECV. Dark count rates of SPDs are respectively 41.7, 18.00, and 15.40Hz when the operation frequency is 1MHz and the detection gate width is 2.5ns. The quantum efficiency of SPDs is 10\% at a wavelength of 1550nm. 

\begin{table}
\centering
\caption{\label{tab:4portexp_insloss}Insertion loss of a 4-port router in the unit of dB.}
\begin{tabular}{ccccc}
	\\
	\hline
    ~      & Port A      & Port B      & Port C      & Port D      \\ \hline
    Port A & ~           & 1.70        & 2.47        & 2.48 \\
    Port B & 2.17        & ~           & 1.64        & 2.74 \\
    Port C & 2.61        & 2.16        & ~           & 2.25 \\
    Port D & 1.96        & 2.66        & 1.99        & ~    \\ \hline
\end{tabular}
\end{table}

With the setup, key broadcast is performed. \textit{Alice} sends a clock signal to three homemade circuits boards. Delay generators on boards delay the single and output delayed signals to trigger laser diodes respectively. To avoid crosstalk, delay times are set different to make laser pulses of different wavelength arrive the quantum router $\text{QR}_1$ at different times. Light laser pulses are transferred though interferometers and modulated with different phases. The pulses are attenuated by mATTs to pseudo single-photon level and multiplexed to a fiber with a WDM. After arriving at the port D of $\text{QR}_1$, pulses are dispatched to different ports and transferred to corresponding users: pulses of 1510nm to \textit{Bob}, pulses of 1530nm to \textit{Charlie} and pulses of 1550nm to \textit{Delta}. Each client uses an interferometer that is identical to the one used by \textit{Alice} to demodulate pulses and detects them with a SPD. The whole process is controlled by two computers, one for the server and the other for three clients. The computers are connected to a local network, which establishes a public communication channel to exchange instructions and key data. At the output of \textit{Alice}, a electronic control optical attenuator eATT is used to attenuate laser pulses and simulate different transmission lengths. 

\begin{figure}[t]
\centerline{\includegraphics[scale=.35]{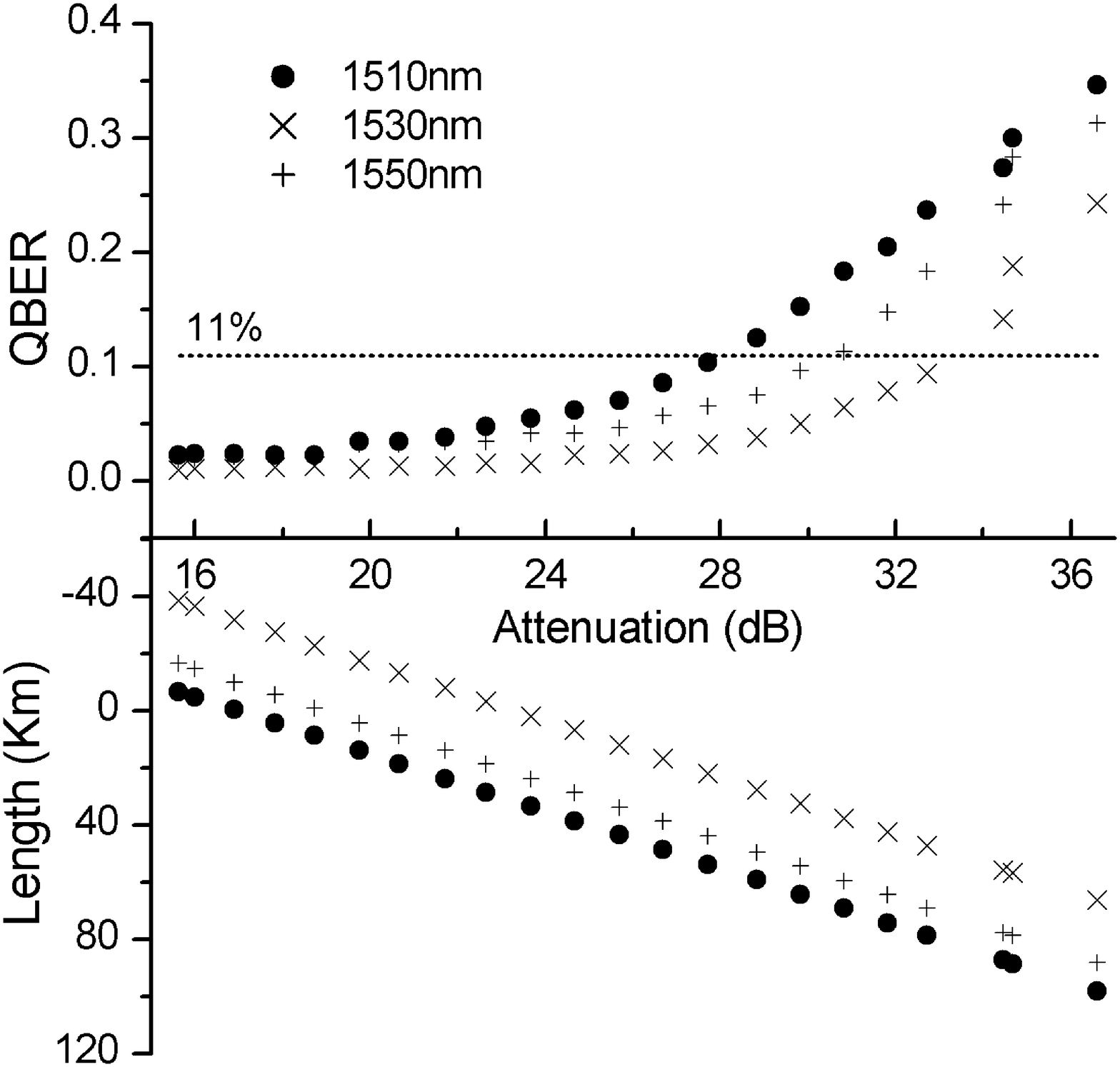}}
\caption{\label{fig:4portexp_rst}The upper figure is quantum bit error rates (QBER) of different wavelengths as function of the attenuation value of eAtt. The lower figure is transmission lengths as function of the attenuation value of eAtt.}
\end{figure}

To synchronize server and clients, 10:90 beam-splitters BS are placed at outputs of each laser diode. Every laser pulse is divided into two parts, the weak one is coupled to a interferometer and the stronger one serves as a synchronous pulse. Three different wavelength synchronous pulses are multiplexed by a WDM and transferred to the quantum router $\text{QR}_2$. This quantum router is identical with the one used for single-photon pulses, so a synchronous pulse and single photon pulse of the same wavelength will transfer to the same client. At each client, synchronous pulses are detected by a InGaAs PIN detector DET. The output of the DET is used to trigger the corresponding SPD.

Figure \ref{fig:4portexp_rst} shows the experiment result. Because three clients operate independently, quantum bit error rates are different each other. With error correction and key reversion operation, four users can share same keys.

In summary, we present a scheme for building quantum key distribution network with wavelength division multiplexing. With the scheme, a network of dozens or hundreds of users is possible. Using server-client model, key unicast, multicast or broadcast is available. Furthermore, a 4-user demonstration network is built up and key exchange is performed.

\medskip
This work was funded by National Natural Science Foundation of China (Grant No. 60121503 and 60537020) and the Knowledge Innovation Project of Chinese Academy of Sciences. Zheng-Fu Han's e-mail address is zfhan@ustc.edu.cn.


\end{document}